\documentclass[prc,nofootinbib,showpacs,twocolumn]{revtex4}

\usepackage{graphicx}
\usepackage[usenames,dvipsnames]{color}
\usepackage{amsmath,amssymb}
\usepackage{hyperref}

\begin{document}
\title{Power spectrum of flow fluctuations in relativistic heavy-ion
collisions}
\author{Saumia P.S.$^1$ \footnote {email: saumia@imsc.res.in}}
\author{Ajit M. Srivastava$^2$ \footnote{email: ajit@iopb.res.in}}
\affiliation{$^1$ The Institute of Mathematical Sciences, Chennai, 
600113, India,\\
$^2$ Institute of Physics, Bhubaneswar 751005, India}

\begin{abstract}
We carry out hydrodynamical simulation of the evolution of fluid
in relativistic heavy-ion collisions with random initial fluctuations.
The time evolution of power spectrum of momentum anisotropies shows 
very strong correspondence with the physics
of cosmic microwave anisotropies as was earlier predicted by us.
In particular our results demonstrate suppression of superhorizon 
fluctuations and the correspondence between the location of the first 
peak in the power spectrum of momentum anisotropies and the length
scale of fluctuations and expected freezeout time scale (more precisely, 
the sound horizon size at freezeout).   
\end{abstract}
\maketitle

It has been recognized for some time now that there are deep 
interconnections between the evolution of flow anisotropies in
relativistic heavy-ion collision experiments (RHICE) and the physics 
of inflationary density fluctuations observed as cosmic microwave 
background radiation (CMBR) anisotropies. Analogies between 
the last scattering surface for CMBR and the freezeout surface in  
RHICE were always mentioned, however these remained at a motivational
level. A rigorous correspondence in the physics of the two systems was
first made by us in ref.\cite{cmbhic} where it was pointed out
for the first time that instead of focusing on first few flow coefficients
(which, until that time, were primarily only even flow coefficients $v_2,
v_4, v_6$) one should plot the entire power spectrum of all flow
coefficients including the odd coefficients. The reason for departure
from conventional focus on only first few even flow coefficients
was the recognition that initial state fluctuations contribute to 
development of all flow coefficients (including the odd ones) even
for central collisions. Many subsequent investigations confirmed
this expectation \cite {flowfluct,srnsn} and indeed now one routinely 
measures odd coefficients (e.g. the triangular flow coefficient $v_3$) and
there have also been several investigations of power spectrum of flow
coefficients upto a large value of $n$ of about 10-12.   

  In this work we present results of hydrodynamic simulations of
the evolution of flow fluctuations and focus on the physics of
various features of the power spectrum keeping in mind the very rich
possibilities of correspondence with the power spectrum of CMBR 
anisotropies. Some of the most important features of CMBR power spectrum
are the presence of acoustic peaks, the relation between the location of
the first peak and the horizon size at last scattering surface, and the
suppression of power in superhorizon modes. These are intimately
connected to the fact the density fluctuations of the universe originate
during an early phase of inflation where causally generated fluctuations
are stretched to superhorizon scales. We study these aspects in 
RHICE using hydrodynamic simulations. We study the evolution of power
spectrum of momentum anisotropies $v_n$ (root mean square values
$v_n^{rms}$ of $v_n$s) and contrast it with the power spectrum of initial 
spatial anisotropies $F_n$ (root mean square values $F_n^{rms}$ of
$F_n$s). We find that for large values of $n$, the two power spectrum 
correlate reasonably well. However for $n$ smaller than a specific value 
(about 3-4 for the cases studied by us) there are significant differences. 
While plot of $F_n^{rms}$ keeps monotonically rising with decreasing $n$, 
the plot of $v_n^{rms}$ starts dropping below $n \sim 3-4$. In \cite{cmbhic} 
we had predicted this using the concept of superhorizon fluctuations,
as spatial anisotropies  larger than a specific wavelength are not 
fully transferred to the momentum anisotropies due to insufficient time
for flow development from pressure gradients over the relevant 
length scales. The first peak therefore contains the information about the 
longest wavelength of fluctuations at the freezeout 
surface. We next consider initial fluctuations of specific wavelengths 
and study the shift of the first peak as the size of initial fluctuations 
is varied. We find that for larger (smaller) initial fluctuations the 
peak shifts to smaller (larger) $n$ roughly by the same scale factor. 
Further, for a fixed initial size of spatial fluctuations, we find that
the first peak in the power spectrum of momentum anisotropies keeps
shifting to lower values of $n$ as time increases. Nature of this shift 
is entirely consistent with the increasing size of the sound horizon with 
time.  The nature and location of the first peak in the power spectrum of 
flow anisotropies thus contains rich information about the initial 
fluctuation spectrum, their evolution, and the freezeout time scale.
We also study time evolution of $v_n$s at different $n$ and show
that oscillations set in for larger values of $n$ early on just
as happens for acoustic oscillations in CMBR power spectrum.
For CMBR power spectrum even the ratios of different peaks contains
important information, e.g. baryon to dark matter ratio. Here, for
the single fluid dynamics we have studied, we find that the ratio of 
first peak to 2nd peak directly relates to the scale of initial 
fluctuations. For multifluid case, with particle species specific
power spectra, different peak ratios may yield important information
for different fluid components of the plasma. 
Our results confirm the expectation that there is a deep 
correspondence between the evolution of fluctuations in RHICE and
that in the early universe. With phenomenal success of CMBR power
spectrum analysis in providing detailed information about the
very early history of the universe, this should provide ample
motivation for detailed investigations of the power spectrum of
flow fluctuations in RHICE for probing the very early stages of
plasma evolution.  

 We have developed two independent codes for hydrodynamic simulations.
One is a 2+1 dimensional code in the framework of Bjorken's
longitudinal scaling expansion model and uses leapfrog algorithm
of 4th order accuracy with a trapezoidal correction\cite{zalesak}. The 
initial conditions for this simulation are generated using a Wood-Saxon 
background plus fluctuations obtained using Glauber Monte Carlo 
\cite{gbmc} where a nucleus-nucleus collision is viewed as a sequence of 
independent binary nucleon-nucleon collisions. The total energy density 
calculated using Glauber optical initial condition for the same parameters is 
distributed equally among the binary collisions as Gaussian fluctuations
(plus the Wood-Saxon part). A lattice equation of state is used for the
evolution of the plasma.

The second code is a full 3+1 dimensional code. This uses leapfrog
algorithm of 2nd order accuracy and uses QGP ideal gas equation of 
state for 2 massless flavors. The flow become unstable for
large velocities. In regions of negligible energy density we
artificially put an upper cutoff (of 0.95) on the fluid speed.
When regions with significant energy density start developing 
instabilities, simulation is stopped. The initial conditions here are
provided in terms of a Wood-Saxon background plus randomly
placed Gaussian fluctuations of specific widths. In order to 
compare with the results with first code, only central rapidity
region is used for calculation of flow fluctuations using transverse
coordinates and momenta. We will present results from this second simulation
as it provides a direct control of spectrum of initial fluctuations.
This way specific patterns observed in the final power spectrum can
be correlated with the properties of initial fluctuations. We have
compared each of these results with the 2+1 dimensional code (by
utilizing partial control over density fluctuations in the Glauber
model in terms of energy density deposition in each binary collision) 
and results of both simulations show similar patterns. 

 We use same methods for calculating spatial and flow anisotropies
as in our earlier work \cite{cmbhic}. We will use $F_n$ to denote Fourier 
coefficients for the spatial anisotropies, and use the conventional 
notation $v_n$ to denote $n_{th}$ Fourier coefficient of the resulting 
momentum anisotropy in ${\delta p}/p$. We do not calculate the average
values of the flow coefficients $v_n$, instead we calculate root-mean
square values of the flow coefficients $v_n^{rms}$. Further, these
calculations are performed in a lab fixed frame, without any
reference to the event planes of different events. Average values of 
$v_n$ are zero due to random orientations of different events. 
As $v_n^{rms}$ will have necessarily non-zero values, physically useful 
information will be contained in the non-trivial shape of the power 
spectrum (i.e. the plot of $v_n^{rms}$ vs. $n$). More precisely, our
focus will be on the structure of the peaks of the plot and evolution
of oscillations in this plot as a function of time. Momentum
anisotropies are calculated by calculating transverse momentum
density at each point from the evolving fluid energy momentum
tensor. Spatial anisotropies are estimated by  calculating the 
anisotropies in the fluctuations in the spatial extent $R(\phi)$
at any given stage, where $R(\phi)$ represents the energy density 
weighted average of the transverse radial coordinate in the angular bin 
at azimuthal coordinate $\phi$. We divide the region in 50 - 100 bins of 
azimuthal angle $\phi$, and calculate the Fourier coefficients $F_n$s of 
the anisotropies in ${\delta R}/R \equiv  (R(\phi) - {\bar R})/{\bar R}$ 
where $\bar R$ is the angular average of $R(\phi)$. Values of root mean 
square values $F_n^{rms}$ of $F_n$s gives us the power spectrum of spatial 
anisotropies. Note that in this way we are representing
all fluctuations essentially in terms of fluctuations in the boundary of
the initial region. Clearly there are density fluctuations in the interior
region as well. However, in order to compare with momentum anisotropies
(which are observed only as a function of azimuthal angle), the full
$x-y$ dependence of spatial inhomogeneities has to be represented
in terms of some average angular fluctuation assuming that the 
representation by fluctuating boundary will capture the essential physics. 
A more careful analysis should include the details of fluctuations in 
the interior regions and then make comparison with momentum anisotropies.
We study evolution starting with initial time scale $\tau_0 = 0.3$ fm
and a temperature $T_0 = $ 500 MeV. Results are presented for 150
events.  Evolution is carried out upto a 
maximum time $\tau = 3.24$ fm. Wood-Saxon background is used
with initial central density corresponding to QGP at $T = T_0$ and
radius of 2.5 fm, with skin thickness of 0.3 - 0.5 fm. For the 3-D code,
Gaussian fluctuations are used with widths varying from 0.25 fm to 1.6 fm.

We first present results relating the location of the first peak to 
the size of initial fluctuations. In CMBR power spectrum, the location of
first peak occurs at around 1$^0$ which corresponds to the the size of
horizon at last scattering surface. More importantly, this is also the
size of largest fluctuation which grows by gravitational collapse, hence
leaving mark of that scale on the peak location in CMBR power spectrum.
We thus expect location of first peak in power spectrum of flow fluctuations
to correlate with the scale of dominant initial fluctuations. Fig.1a shows 
the flow fluctuations power spectrum  at $\tau = 1.98$ fm for different
sizes of initial fluctuations. We note the net shift in the peak position 
to smaller values of $n$ with increasing size of initial 
fluctuations. This clearly shows that the peak position directly probes 
length scale of relevant fluctuations.

We next study the prediction made in \cite{cmbhic} that large scale
spatial anisotropies will not be fully reflected in momentum anisotropies.
More precisely, spatial fluctuations with wavelengths larger than sound 
horizon at freezeout should show suppressed flow fluctuations, this was 
termed as superhorizon suppression in \cite{cmbhic}. 
Fig.1b shows the plots of initial power spectrum of spatial anisotropies
for initial fluctuations with sizes as in Fig.1a. Comparison of
Fig.1a and Fig.1b shows important qualitative difference for low $n$.
We note that for $n$ larger than about 4, both plots show similar
pattern. However for smaller $n$, the two plots show
dramatic difference. Plot of $F_n^{rms}$ keeps rising monotonically
with decreasing $n$. However, plot of $v_n^{rms}$ shows a drop for
low $n$ values. As predicted in ref.\cite{cmbhic} this is exactly what
one expects from the suppression of those spatial modes which do not 
get enough time to transfer to momentum anisotropies, that is due
to their superhorizon nature. It is important to note that heavy-ion
data has consistently shown this drop for low $n$ values (first
plot was shown by Sorenson \cite{srnsn}, who had associated this
drop with the prediction of superhorizon suppression as discussed
in \cite{cmbhic}. See \cite{peak} for other results on this).  

\begin{figure}[!htp]
\begin{center}
\includegraphics[width=0.35\textwidth]{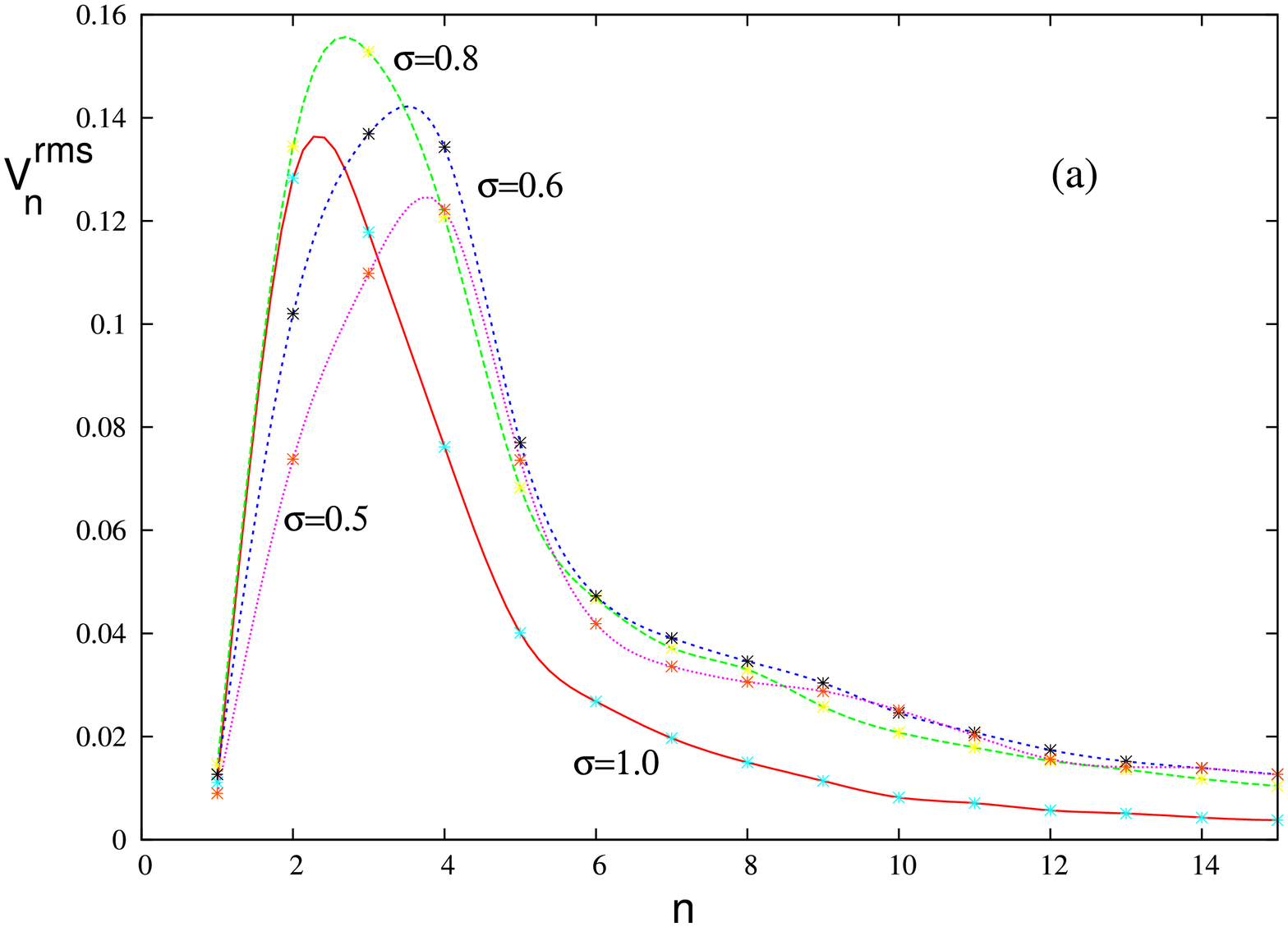}
\includegraphics[width=0.35\textwidth]{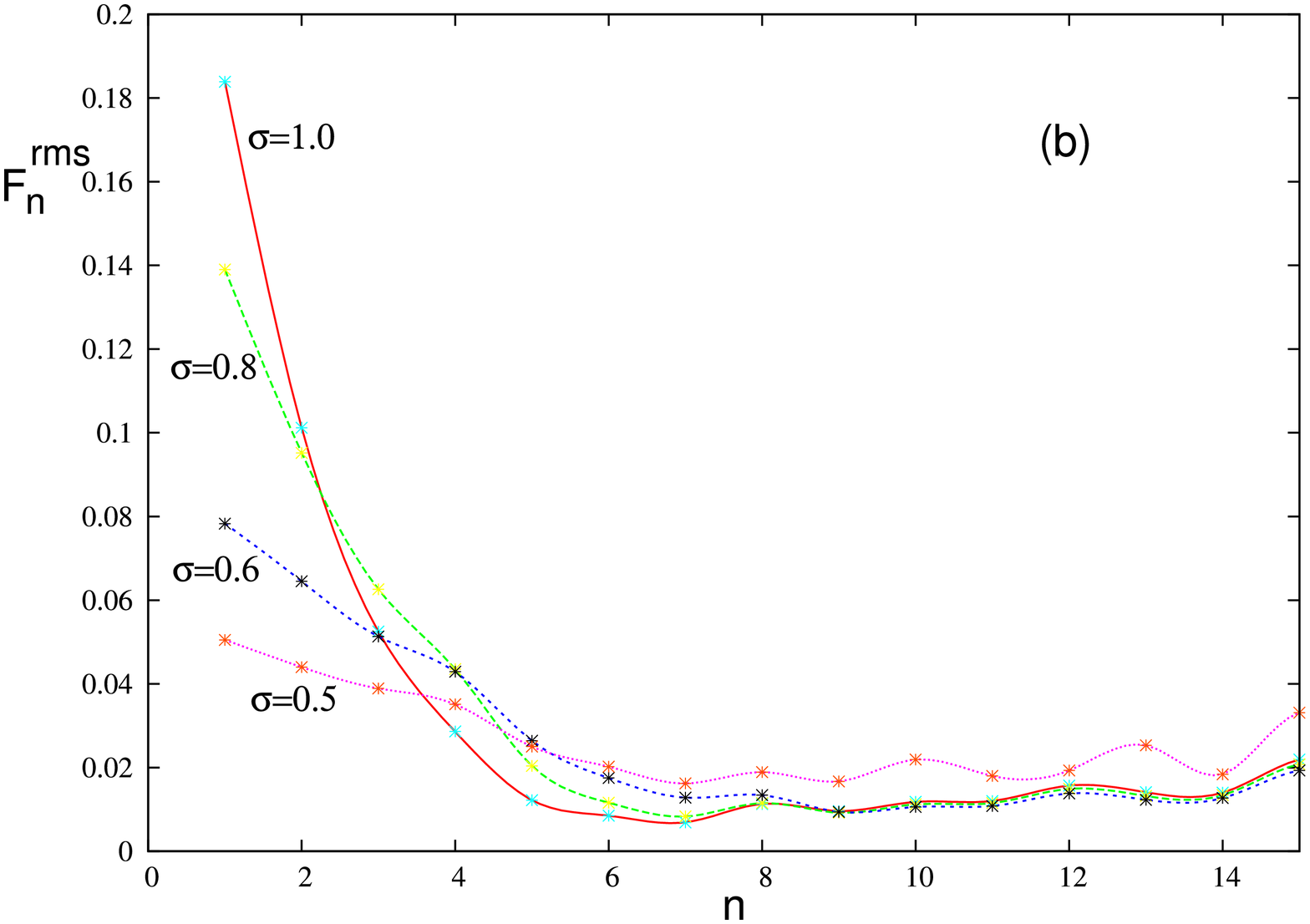}
\caption{(a) Plots of $v_n^{rms}$ for different sizes $\sigma$ (in fm)
of initial fluctuations at $\tau = 1.98$ fm. Note the net 
shift in the peak position to smaller values 
of $n$ with increasing size of initial fluctuations. (b) Plots of 
initial power spectrum $F_n^{rms}$ of spatial anisotropies $F_n$. 
Plots of $F_n^{rms}$ in (b) keep rising monotonically with decreasing $n$. 
However, plots of $v_n^{rms}$ in (a) show a drop for low $n$ values.}
\label{fig1}
\end{center}
\end{figure}

 As we mentioned, the location of first peak in CMBR case directly relates to
the size of sound horizon at surface of last scattering. We show in Fig.2,
plots of $v_n^{rms}$ for different times for one specific simulation
(here with initial fluctuation size $\sigma = $ 0.6 fm). Note that there
is a consistent shift in the location of first peak towards lower values of 
$n$ as time increases. If the plasma undergoes freezeout at some specific 
time $\tau_{fr}$, then the final power spectrum of hadrons will be
the one corresponding to the power spectrum shown in Fig.2 at
$\tau = \tau_{fr}$. Earlier freezeout (with small values of $\tau_{fr}$)
will lead to first peak at larger values of $n$ while the peak will
shift to lower $n$ if $\tau_{fr}$ is larger. Thus the location of
first peak in the power spectrum of flow fluctuations directly gives
information about the freezeout time scale in RHICE.  

\begin{figure}[!htp]
\begin{center}
\includegraphics[width=0.35\textwidth]{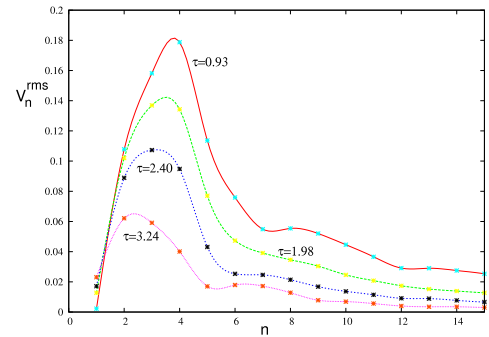}
\caption{Plots of $v_n^{rms}$ at different times $\tau$ (in fm) for one 
specific simulation (here with initial fluctuation size $\sigma = $ 
0.6 fm). Note consistent shift of the location of first peak towards 
lower values of $n$ as time increases.}
\label{fig2}
\end{center}
\end{figure}

 We now study the detailed evolution of plots of $v_n^{rms}$ in time and
focus on the development of acoustic oscillations. It is well known
that for elliptic flow the momentum anisotropy starts decreasing after 
some time (due to decrease in spatial anisotropy in time), but it never
reverses sign. This is because the expected oscillation time scale for
elliptic flow is too large and radial flow starts dominating before that.
It was pointed out in \cite{cmbhic} that one should expect oscillations to
develop fully for shorter wavelength modes, just as for CMBR where
modes of short wavelengths undergo several oscillations by the time
the largest mode starts entering horizon. Fig. 3 shows the time evolution
of $v_n^{rms}$ for specific values of $n = 7, 9, 10$, and 11. We note
that oscillations develop early for larger values  of $n$ (= 10,11)
compared to the case for $n = 7, 9$). Again, this
demonstrates the detailed correspondence in the development of
acoustic peak structure for flow fluctuations just as in the CMBR case.

\begin{figure}[!htp]
\begin{center}
\includegraphics[width=0.35\textwidth]{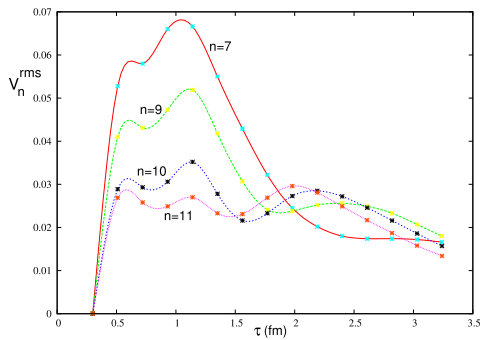}
\caption{Time evolution of $v_n^{rms}$ for specific values of $n$.
Note that oscillations develop early for larger values  of $n$ 
(= 10,11) compared to the case for $n = 7, 9$).}
\label{fig3}
\end{center}
\end{figure}

 For CMBR case, peaks at larger values of ${\it l}$ correspond to regions 
which have undergone several oscillations. Similar physics is seen here
in Fig.4a which shows varying heights of 1st and 2nd peaks (at the
final stage $\tau = 3.24$ fm) depending on initial fluctuation size.
For CMBR, ratio of second peak to the first peak is directly related to 
the baryon matter to dark matter ratio. For single fluid component case
studied here, we find this ratio to directly relate to the size of the 
initial fluctuations (with smaller fluctuations having smaller 1st peak to 
2nd peak height ratio). Fig.4b shows the plot of this ratio as a function 
of $\sigma$, the size of initial fluctuations. (More study is needed to 
see specific dependence of this ratio on $\sigma$.) When one includes 
aspects of multifluid component nature of QGP, with different particle 
species interacting with each other, it is entirely possible that a 
species specific plot of $v_n^{rms}$ may encode rich information about 
different fluid components in terms of ratios of different peak heights.

\begin{figure}[!htp]
\begin{center}
\includegraphics[width=0.23\textwidth]{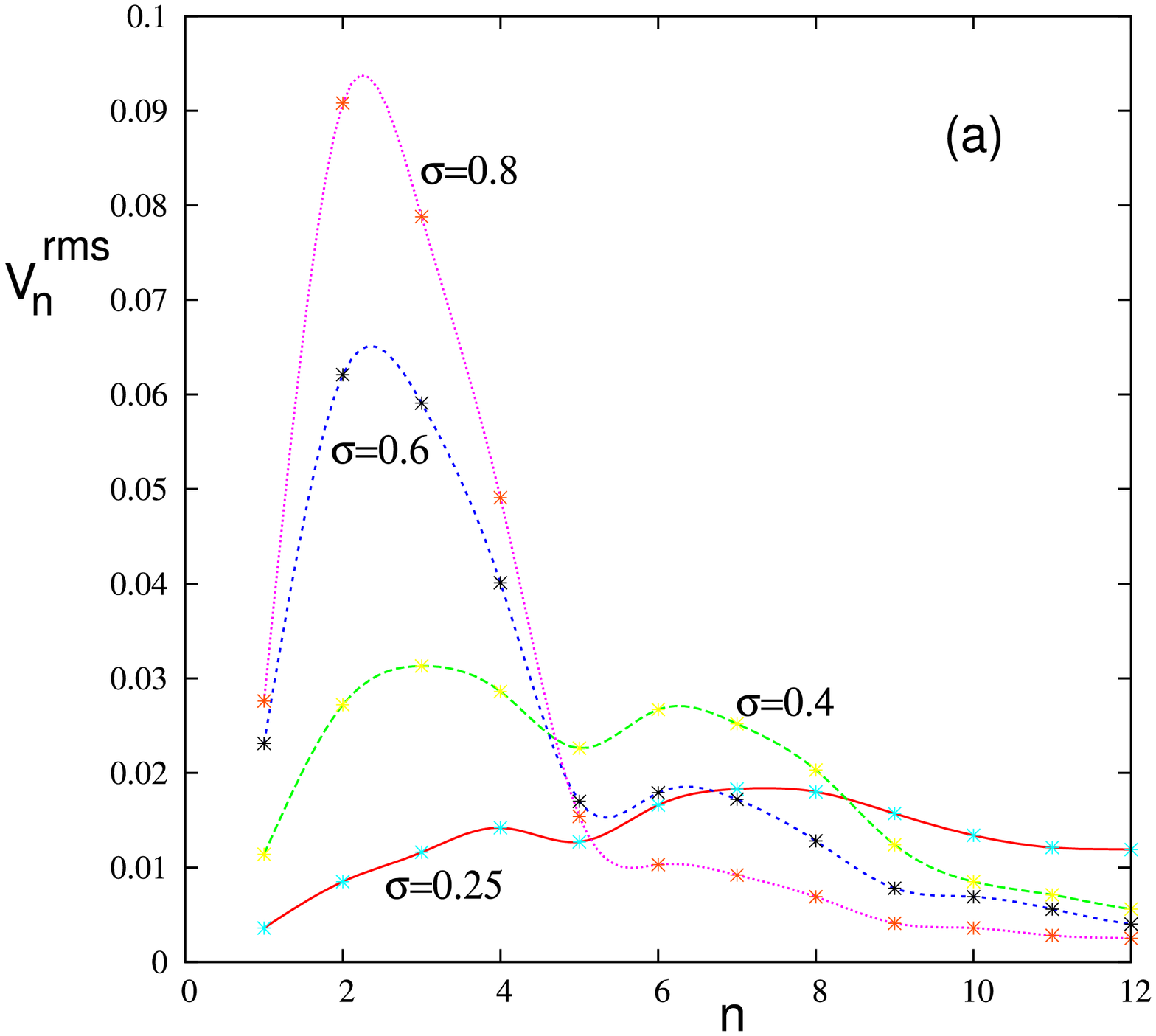}
\includegraphics[width=0.23\textwidth]{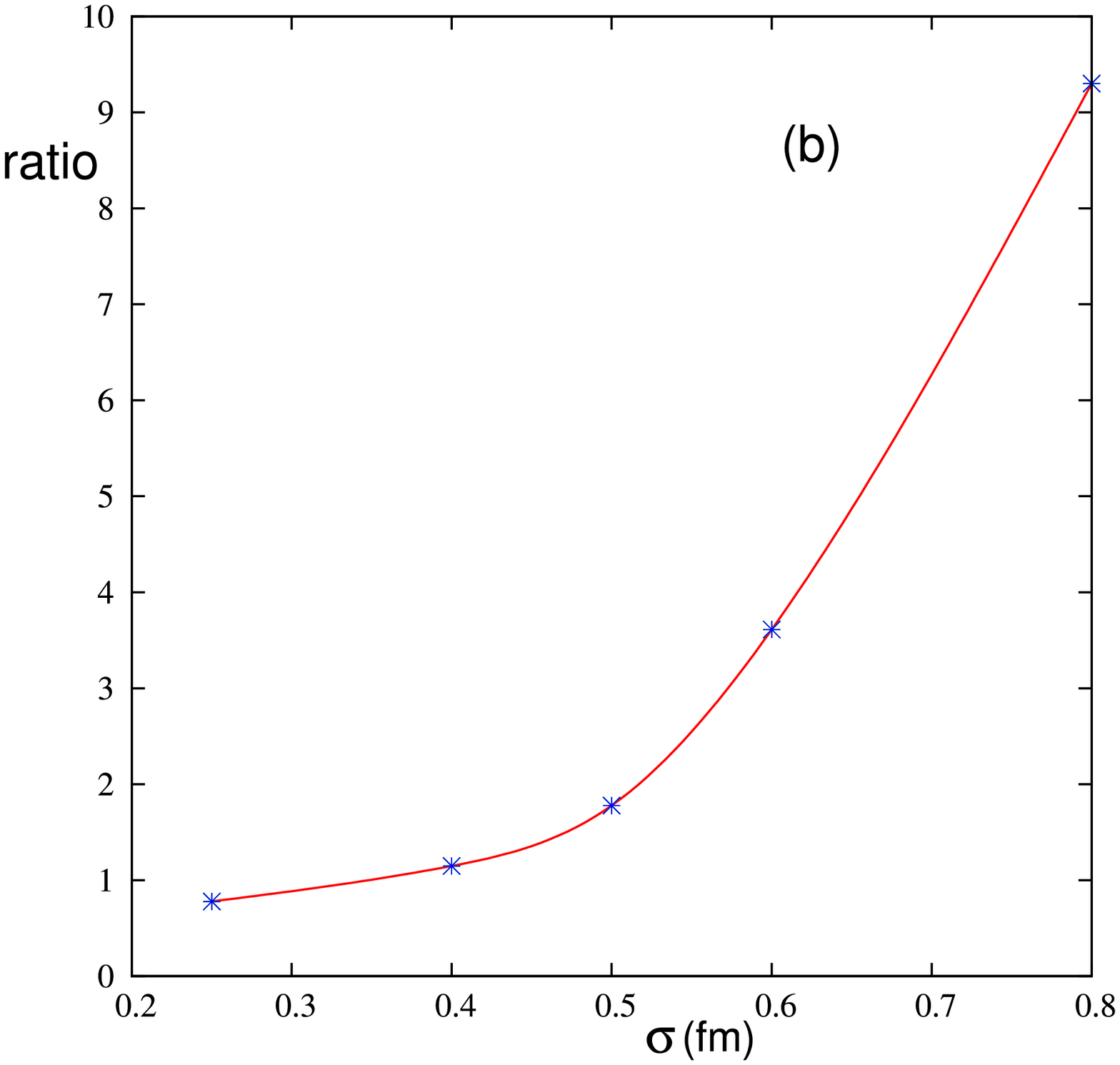}
\caption{(a) Different plots show varying heights of 1st and 2nd peaks 
(at the final stage $\tau = 3.24$ fm) depending on initial fluctuation 
size $\sigma$ (in fm). (b) Plot of the ratio of the heights of the 1st 
peak to the 2nd peak in (a) as a function of $\sigma$}
\label{fig4}
\end{center}
\end{figure}

 We conclude by emphasizing that our hydrodynamic simulation results
have confirmed all the qualitative features predicted in \cite{cmbhic}
for the presence of acoustic peaks in the power spectrum of flow
fluctuations just as for CMBR peaks. One important feature remaining
is the coherence of fluctuations. Coherence in CMBR case results in
pronounced peak structure. However, we have not been able to analyze
the nature of dissipation, and contributions of statistical fluctuations in
our simulations to a level to comment on whether any possible coherence
will be visible in the multiple peak structure of the power spectrum
at RHICE. Nonetheless, as mentioned above, the detailed nature of the 
first peak itself has very rich information about initial fluctuations. 
A details analysis of the second peak (and other peaks as well) can give
immense amount of information about the plasma evolution, including
additional constraints on viscosity and equation of state etc. Power
spectrum of different particle species (baryons vs. mesons, non-strange
vs. strange particles etc.) will give information about interactions
between different particle species and their respective contributions to
plasma evolution, just as different peak ratios for CMBR give information
about baryon to dark matter ratio etc. Another line of deep correspondence
with CMBR physics was initiated by us in \cite{rhicmag} where
effects of early stage magnetic fields on power spectrum of flow 
fluctuations was studies. Magnetohydrodynamics simulations are needed
for detailed studies of it. We hope to present such results in future.

 We are very grateful to Sanatan Digal for very helpful discussions,
especially for simulations. We also thank Partha Bagchi, Arpan Das, 
Shreyansh S. Dave, Pranati Rath, and Somnath De for useful discussions. 



\begin{thebibliography}{99}

\bibitem{cmbhic}  A. P. Mishra, R. K. Mohapatra, P. S. Saumia, and 
A. M. Srivastava, Phys. Rev. {\bf C 77}, 064902 (2008); 
Phys. Rev. {\bf C 81}, 034903 (2010).

\bibitem{flowfluct} P. Sorenson, J.Phys. G37 (2010) 094011; 
B. Alver and G. Roland, Phys. Rev. {C 81} (2010)
054905; A. Mocsy and P. Sorenson, Nucl. Phys. {\bf A855} 
(2011) 241, J.I. Kapusta, Nucl. Phys. {A862-863} (2011) 47;
J.Y. Ollitrault, J. Phys. Conf. Ser. 312 (2011) 012002.

\bibitem{srnsn} P. Sorenson, arXiv:0808.0503.

\bibitem{zalesak} Steven T. Zalesak, J. Comput. Phys. {\bf 31}, 335 (1979).

\bibitem{gbmc} Michael L. Miller, Klaus Reygers, Sthephen J. Sanders 
and Peter Steinberg, Ann. Rev. Nucl. Part. Sci. {\bf 57}, 205 (2007) 

\bibitem{peak} S. Mohapatra, Journal of Physics: Conference Series
389 (2012) 012011; P. Staig and E. Shuryak, J.Phys. G38 (2011) 124039;
T. Gorda and P. Romatschke, Phys.Rev. C90 (2014) 5, 054908. 

\bibitem{rhicmag} R. K. Mohapatra, P. S. Saumia, and A. M. Srivastava,
Mod. Phys. Lett. {\bf A 26}, 2477 (2011).

\end{thebibliography}
\end{document}